\documentclass[aps,prb,twocolumn,showpacs]{revtex4}
\usepackage{graphicx}

\begin{document}

\title{Magnetism of ordered Sm/Co(0001) surface structures} 

\author{J.E. Prieto$^1$}
 \email{joseemilio.prieto@uam.es}
 \altaffiliation[present address: ] {Centro de Microan\'alisis de Materiales, Universidad Aut\'onoma de Madrid, E-28049 Madrid, Spain}
\author{O. Krupin$^1$}
\author{S. Gorovikov$^2$}
\author{K. D\"obrich$^1$}
\author{G. Kaindl$^1$}
\author{K. Starke$^1$}
\affiliation{$^1$Institut f\"ur Experimentalphysik, Freie Universit\"at 
Berlin, Arnimallee 14, D-14195 Berlin, Germany\\
$^2$MAX-Lab, Lund University, P.O. Box 118, Lund S-22100, Sweden}


\date{\today}

\begin{abstract}
The epitaxial system Sm/Co(0001) was studied for Sm coverages up to
1 monolayer (ML) on top of ultrathin Co/W(110) epitaxial films.
Two ordered phases were found for 1/3 and 1 ML Sm, respectively.
The valence state of Sm was determined by means of photoemission
and magnetic properties were measured by magneto-optical Kerr effect.
We find that 1 ML Sm causes a strong increase of the coercivity with
respect to that of the underlying Co film.
Element-specific hysteresis loops, measured by using resonant soft x-ray
reflectivity, show the same magnetic behaviour for the two elements.
\end{abstract}


\maketitle

\section{Introduction}
\label{intro}

Magnetic metals can be classified into two main groups. On the one hand, the
transition metals (TMs), where the magnetic moments are carried by the partly
{\em itinerant}, strongly overlapping 3$d$ electrons. Due to strong
crystal-electric fields, the orbital moments are mostly quenched, and the
magnetic moments have predominantly spin character. The magnetic coupling
is therefore strong, giving rise to ordering temperatures as high as
$\approx$~1000 Kelvin, while the magnetic anisotropies are relatively small.
On the other hand, rare-earth (RE) magnetism is determined by the
{\em localized}, atomic-like character of the magnetic moments
of the 4$f$ shell, which in general contain both a spin and an orbital part.
Non-vanishing orbital moments give rise to non-spherical 4$f$ charge
distributions that lead to strong ``single-ion" contributions to the
magnetic anisotropy.
The very small overlap between the 4$f$ orbitals of neighbouring atoms
is responsible for a negligible direct exchange interaction
between the 4$f$ moments of RE ions. Instead, they couple indirectly
through the conduction electrons (RKKY interaction), a mechanism that
leads to ordering temperatures typically lower than room temperature
(RT) in RE metals.

Some intermetallic compounds containing both RE and TM ions combine the
magnetic properties of the two classes of components. For example,
the Co-Sm and Nd-Fe-B systems include the magnetically hardest materials
known today.
In these compounds, the high magnetic anisotropies are induced by
the RE ions, while the characteristic high ordering temperatures
of the ferromagnetic TMs are retained\cite{mhh93}.

The trend in magnetic storage technology towards ever higher densities
requires to reduce system dimensions to a degree where the superparamagnetic
limit is approached~\cite{wm99}.
A possible solution is the development of thin films of materials
with high magnetic anisotropy energies per unit volume that
could retain high ordering temperatures and high coercivities
at RT even when system sizes approach the nanometer scale.
A promising material is Co$_5$Sm, in which a relatively small fraction of
the RE metal Sm renders the material much harder than pure
Co~\cite{herbst91,buschow91}.
Hence, it is interesting to study the effect of Sm on the magnetic properties
of very thin Co films.

Here, we report on a study of the epitaxial system Sm/Co(0001)
on W(110) with Sm coverages up to 1~ML, where we found
several ordered surface phases. Their magnetic properties were
studied by means of visible-light and soft x-ray magneto-optical Kerr effect
(MOKE), and their electronic structure was investigated by photoelectron
spectroscopy.

\section{Experimental}
\label{exp}

Co films of about 10~ML thickness were prepared
by metal vapour deposition in ultra-high vacuum (UHV)
on a W(110) single-crystal substrate.
To this end, a high-purity Co rod was heated by electron bombardment.
Sm was deposited from a W crucible.
The same substrate and evaporators were used in all experiments.
Deposition rates were of the order of 1~ML per minute.
The crystallinity of the surfaces was checked by
low-energy electron diffraction (LEED) using a rear-view optics.
MOKE hysteresis loops were recorded {\em in situ} employing a rotatable
electromagnet with a soft-iron yoke\cite{magnet}, with external magnetic
fields up to 2~kOe applied in-plane along the substrate
bcc[$\overline{1}$10] direction; this corresponds to the easy axis of
magnetization of the thin epitaxial Co/W(110) films\cite{fkg95}.
Resonant soft x-ray reflectivities using circularly-polarized (CP) light
were measured for films prepared {\em in situ} in the same UHV chamber 
attached to the UE52-SGM undulator
beamline of BESSY II. The specularly reflected intensity was detected
by a Si photodiode mounted on a home-made $\theta-2\theta$ goniometer
inside the UHV chamber.
PE experiments were performed on films prepared in the same way at the
I-311 undulator beamline of MAX-Lab in Lund, Sweden, which is equipped
with a display-type electron analyzer. Spectra shown here were measured
at normal emission in the angle-integrated mode, with an acceptance angle
of $\pm12.5^\circ$.

\section{Results and discussion}
\label{r&d}

\begin{figure}
\begin{center}
\includegraphics*[width=8.5cm]{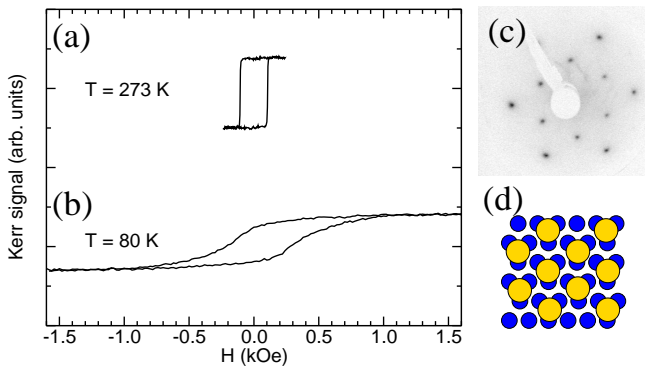}
\caption{Visible-light MOKE hysteresis curves of
Sm(1/3~ML)/Co(8~ML)/(0001) measured at (a)
273~K and (b) 80~K. The LEED pattern for an electron energy of 150~eV,
shown in (c) with inverted contrast, corresponds to a 
$(\sqrt{3}\times\sqrt{3}) {\rm R} 30^\circ$ superstructure. 
In (d), the proposed atomic structure
of this phase (top view) is shown schematically; large and small
circles represent Sm and Co atoms, respectively.}
\label{sqrt3}
\end{center}
\end{figure}

\begin{figure}
\begin{center}
\includegraphics*[width=8.5cm]{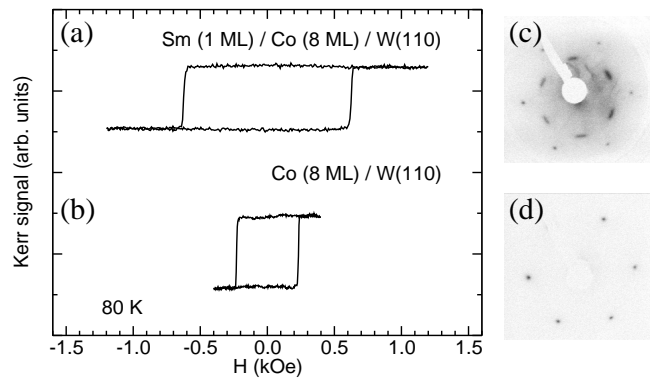}
\caption{(a) Visible-light MOKE hysteresis curve of
Sm(1~ML)/Co(8~ML)/W(110) measured at 80~K. In (b), the corresponding curve
measured under the same conditions for the 8~ML Co/W(110) film prior
to Sm deposition is shown. LEED patterns (150~eV, inverted contrast)
of the two surfaces are displayed in (c) and (d), respectively.}
\label{sml}
\end{center}
\end{figure}

Epitaxial Co films on W(110) with thicknesses larger than about 5~ML
show a (1$\times$1) LEED pattern of hexagonal symmetry
(see, e.g. Fig.~\ref{sml}d), which reflects
the fact that the hexagonal base planes of Co(0001) are parallel to W(110).
In agreement with previous findings~\cite{fkg95,jbg89,osg90,kb93},
the growth proceeds in the Nishiyama-Wasserman orientation, i.e.,
the close-packed Co rows along [11$\overline{2}$0] run parallel to W[001].
In this thickness range, the epitaxial strain amounts to a few
percent~\cite{fkg95}.
Upon deposition of 1/3~ML Sm, a
$(\sqrt{3}\times\sqrt{3}) {\rm R} 30^\circ$ superstructure
appears in the LEED pattern, as shown in Fig.~\ref{sqrt3}c.
The magnetization curves measured by visible-light MOKE on this surface
are also shown in Fig.~\ref{sqrt3}.
At 273~K, the hysteresis loop has a square shape and the coercivity
amounts to 100~Oe. Upon cooling down to 80~K, the coercivity increases to
250~Oe and the shape of the hysteresis becomes more elongated.

For higher Sm coverages, between 2/3~ML and 1~ML, a different LEED pattern
appears. It is shown in Fig.~\ref{sml}c, in comparison with the hexagonal
pattern of the clean Co film (Fig.~\ref{sml}d).
The superstructure spots in Fig.~\ref{sml}c appear close to those of the
$(\sqrt{3}\times\sqrt{3}) {\rm R} 30^\circ$ structure observed for lower
coverages, but they now have an elongated shape along the tangential
direction.
The hysteresis curve measured for this phase at 80~K is shown in
Fig.~\ref{sml}a.
Compared with that of the clean Co film, the coercivity has increased
from 230 to 630~Oe, i.e. by a factor of about 3. This effect allows to
consider the Sm/Co/W(110) system as a prototype of a TM film with increased
anisotropy due to the deposition of a small amount of a RE metal.

\begin{figure}
\begin{center}
\includegraphics*[width=7.5cm]{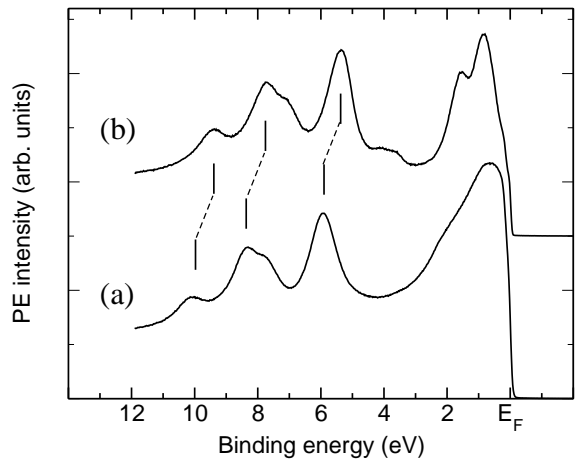}
\caption{Normal-emission photoemission spectra of
(a) Sm(1/3~ML)/Co/W(110) (showing the 
$\sqrt{3}\times\sqrt{3}) {\rm R} 30^\circ$ superstructure) and (b)
Sm(1~ML)/Co/W(110), both recorded with 141-eV photons. The shifts of the
4$f^5$ multiplet lines are indicated by vertical bars.}
\label{pe}
\end{center}
\end{figure}

In Sm compounds, the electronic structure is strongly influenced by the
valence state of the Sm ion. This is of particular importance for surface
phases, because pure Sm metal is known to be trivalent in the bulk but
divalent at the surface layer~\cite{wc78}. 
The reason is that the energy cost of promoting an electron from the 4$f$
shell to the (6$s$5$d$) valence band is not compensated at the
low-coordinated surface layer by stronger bonding. In order to determine
the valence state of Sm, we performed PE experiments on both epitaxial
Sm/Co(0001) phases using a photon energy of 141~eV to resonantly
enhance the Sm features (4$d$-4$f$ resonance).
Figure~\ref{pe} shows the valence-band PE spectra including the Sm
4$f$-multiplet structure. Both phases show strong emission in the
region extending from the Fermi level to approximately 2~eV binding energy.
This is caused by the partially filled 3$d$-band of Co. Furthermore,
the characteristic set of peaks in the binding-energy region from
5 to 10~eV correspond to the final-state $4f$ multiplet reached from trivalent
($4f^5$) Sm~\cite{wc78}. 

The PE spectrum of the $(\sqrt{3}\times\sqrt{3}) {\rm R} 30^\circ$ phase
contains only features characteristic for trivalent Sm~\cite{wc78,sbn89},
i.e., Sm PE peaks at binding energies of 5.9, 8.3 and 10.0~eV.
On the other hand, the 4$f$ multiplet structure found for higher
Sm coverages is shifted by 0.6~eV towards the Fermi level, so that
the PE peaks appear at 5.3, 7.7 and 9.4~eV. In addition, the spectrum
of Fig.~\ref{pe}b shows three features closer to the Fermi level,
at binding energies of 0.8, 1.6 and 3.9~eV, respectively, which can be
assigned to divalent ($4f^6$) Sm ions~\cite{lb79,sbn89}.

We interpret the $(\sqrt{3}\times\sqrt{3}) {\rm R} 30^\circ$ superstructure
in terms of the formation of a magnetic Sm/Co surface phase. Based on the
symmetry of the diffraction pattern as well as the known amount of deposited
Sm (1/3~ML), we propose the model for the atomic arrangement in this phase 
that is shown in Fig.~\ref{sqrt3}d; the Sm atoms occupy 3-fold coordinated 
sites on the topmost Co layer. From our analysis we cannot conclude which
of the two inequivalent adsorption sites on the hexagonal close-packed surface
layer ({\em fcc} or {\em hcp}) is preferred. The PE spectra in Fig.~\ref{pe}a
show that the Sm ions are trivalent in this phase.
The LEED pattern in Fig.~\ref{sml}c indicates for the Sm(1~ML) phase a 
similar structure as for the Sm(1/3~ML) phase showing the 
$(\sqrt{3}\times\sqrt{3}) {\rm R} 30^\circ$ superstructure,
yet with some degree of rotational disorder.
The 1-ML phase contains both trivalent and divalent Sm ions, as shown
by the PE spectrum of Fig.~\ref{pe}b. This points towards the presence
of ``interface" and ``surface" Sm atoms, assuming that further
deposition of Sm on top of the $(\sqrt{3}\times\sqrt{3}) {\rm R} 30^\circ$
surface does not significantly distort the proposed structure.

The magnetic hysteresis curves of
Sm(1/3~ML)/Co/W(110) at 80~K (Fig.~\ref{sqrt3}b)
show a more complex shape than that of the Sm(1~ML) phase at the same
temperature (Fig.~\ref{sml}a).
Besides lower coercivity, it reveals a reduced remanence.
This may be due to a partial reorientation of the magnetization of the film
at lower temperatures. The rectangular hysteresis of the 1-ML phase displayed
in Fig.~\ref{sml} shows again a simple in-plane magnetization loop.

The enhancement of the coercivity of the Co film by Sm can be qualitatively
understood in terms of the Sm single-ion anisotropy~\cite{gls79}.
The aspherical Sm 4$f$ charge distribution is sensitive to the crystal
field particularly at sites of reduced symmetry like at the surface.
The temperature dependence of the magnetic behavior may be related to a
mixing of the multiplet states $J$=7/2 and $J$=9/2 with the ground-state
multiplet state $J$=5/2 of trivalent Sm due to the combined action of
crystalline-electric and exchange fields~\cite{bdw74}.

\begin{figure}[ht]
\begin{center}
\includegraphics*[width=7.5cm]{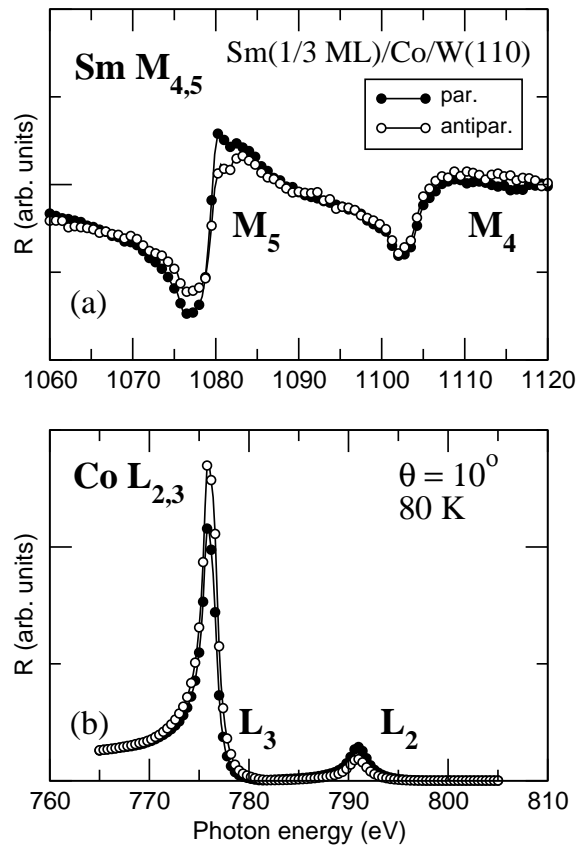}
\caption{Dichroic x-ray reflectivity spectra recorded across (a) the
Sm $M_{4,5}$ and (b) the Co $L_{2,3}$ resonances of
Sm(1/3~ML)/Co(10~ML)/W(110).  
Closed and open circles correspond to nearly parallel and antiparallel 
orientations of photon spin and sample magnetization, respectively.}
\label{rf1}
\end{center}
\end{figure}

\begin{figure}[ht]
\begin{center}
\includegraphics*[width=7.5cm]{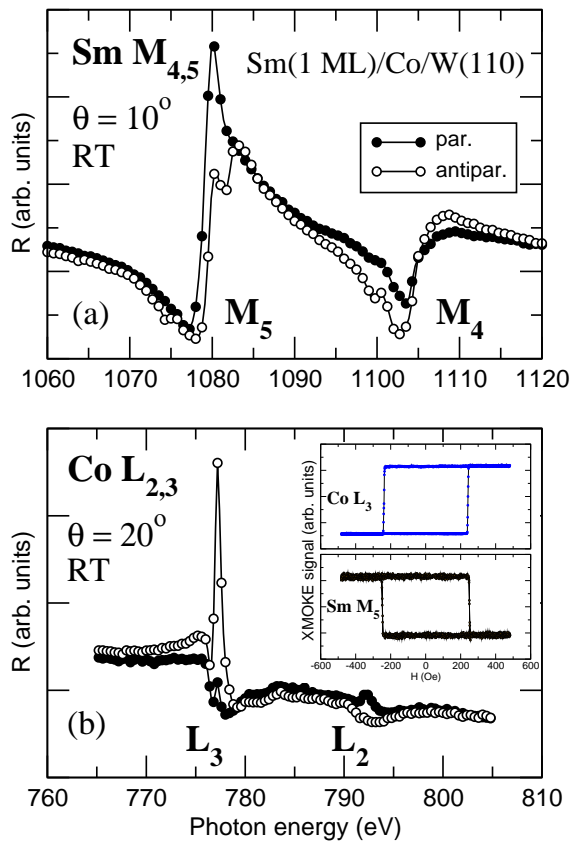}
\caption{Dichroic x-ray reflectivity spectra recorded across (a) the
Sm $M_{4,5}$ and (b) the Co $L_{2,3}$ resonances of
Sm(1~ML)/Co(10~ML)/W(110). Closed and open circles correspond to nearly
parallel and antiparallel orientations of photon spin and sample 
magnetization, respectively. 
The insert shows element-specific hysteresis loops measured at the 
Sm $M_5$ and Co $L_3$ maxima with an x-ray incidence angle of 10$^\circ$.}
\label{rf2}
\end{center}
\end{figure}

The presence of two magnetic elements, Sm and Co, raises the issue of their
possibly different magnetic behaviour. In order to address this point,
we performed element-specific magnetization measurements using
resonant soft x-ray scattering at elemental absorption thresholds.
Figures~\ref{rf1} and ~\ref{rf2} show soft x-ray reflectivity spectra 
measured on both of the studied Sm/Co(0001) ordered structures.
The samples were remanently magnetized in-plane, and circularly polarized
(CP) light was used with the photon spin almost parallel or antiparallel
to the sample magnetization direction.
Figure~\ref{rf1} corresponds to 
Sm(1/3~ML)/Co/W(110), 
Fig.~\ref{rf2} to Sm(1~ML)/Co/W(110). 
The spectra recorded across the Sm $M_{4,5}$
and Co $L_{2,3}$ thresholds are shown in the top (a) and bottom (b) panels
of both figures, respectively.

The striking differences between the dichroic Co $L_{2,3}$ reflectivity spectra
displayed in Figs.~\ref{rf1}b and~\ref{rf2}b are mainly due to the different
angles of x-ray incidence (10$^\circ$ and 20$^\circ$, respectively), although
the samples differ in Sm coverage and temperature as well.
Similarly drastic changes in Co $L_{2,3}$ specular reflectivity spectra
with incidence angle have previously been observed.~\cite{cck94}
They originate from interference, as the soft x-ray wavelenght around
the Co $L_{2,3}$ edge is comparable to the Co film thicknesses in the 
nanometer scale.
The photon-energy range extending to some 10 eV below the resonance maxima
is particulary sensitive to interference, favoured by the long 
x-ray penetration length due to reduced absorption
(small imaginary part of the refractive index $n$) and to the absence of
total internal reflection (real part of $n$ larger than 1).~\cite{phk03,pkd05}

The spectra of both structures contain magnetic contrast, allowing to
perform XMOKE measurements at the Sm $M_5$ and Co $L_3$ thresholds. 
XMOKE curves for the Sm(1~ML)/Co/W(110) phase recorded at an incidence 
angle of 10$^\circ$ are shown in the insert of Fig.~\ref{rf2}(b).
The reduced coercivity ($H_c \approx$~250~Oe) as compared to similar films 
displayed in Fig.~\ref{sml}a is mainly due to the different temperature; 
the slightly different Co thickness is known to play a minor role 
in this range.~\cite{cmh01}
The element-specific hysteresis loops
of both elements reveal the same coercivity,
showing that the film magnetization reverses simultaneously at the
Sm/Co interface and deeper inside the Co film.
Sm/Co films of different thicknesses showed always the same magnetic
behaviour for the two elements.

Summarizing, we have found and characterized two ordered phases in the
Sm/Co system. The Sm(1~ML)/Co phase shows an increased coercivity
by a factor of 3 with respect to a pure Co film of the same thickness.
Further experiments aiming at a detailed structural and morphological
characterization of these phases are under way.

\acknowledgments

J.E. Prieto gratefully acknowledges financial support from the
Alexander-von-Humboldt Stiftung and the Spanish MEC (EX2001).
The authors thank F. Senf for experimental assistance at BESSY.
This work was financed by the German BMBF (05 KS1 KEC/2).

\bibliography{smco}

\end{document}